\documentclass[11pt]{article}
\usepackage{moriond,epsfig}

\def\Journal#1#2#3#4{{#1} {\bf #2}, #3 (#4)}

\def\bea{\begin{eqnarray}}
\def\eea{\end{eqnarray}}

\newcommand{\be}{\begin{equation}}
\newcommand{\ee}{\end{equation}}
\def\bea{\begin{eqnarray}}
\def\eea{\end{eqnarray}}

\newcommand{\gr}{\kern 2pt\hbox{}^\circ{\kern -2pt K}} %  ====> GRADI KELVIN
\newcommand{\oml}{\Omega_{\Lambda}}
\newcommand{\omm}{\Omega_{m}}
\newcommand{\brr}{\begin{array}}
\newcommand{\err}{\end{array}}
\newcommand{\ltsima}{$\; \buildrel < \over \sim \;$}
\newcommand{\simlt}{\lower.5ex\hbox{\ltsima}}
\newcommand{\gtsima}{$\; \buildrel > \over \sim \;$}
\newcommand{\simgt}{\lower.5ex\hbox{\gtsima}}
\begin{document}
\vspace*{4cm}
\title{ IRON ABUNDANCES IN HYDRODYNAMICAL SIMULATIONS OF GALAXY CLUSTERS}

\author{ R. Valdarnini }

\address{SISSA, Via Beirut 2-4, 34014 Trieste, Italy}

\maketitle\abstracts{
Hydrodynamical SPH simulations of galaxy clusters are used to investigate the
metal enrichment of the intracluster medium. The final metallicity abundances
of the simulated clusters are determined according
to the numerical resolution and a number of model parameters.
For a fiducial set of model prescriptions the results of the simulations
indicate iron abundances  in broad agreement with data.
Final X-ray properties are not sensitive to the heating of the ICM.
This supports a scenario where the ICM evolution of cool clusters is driven 
by radiative cooling.
}

\section{Introduction}

 A large set of observations shows that the ICM of galaxy clusters is 
rich in metals ( see ~\cite{fu20,ma20} and references cited therein).
Abundance gradients have also been measured 
\cite{ez97,dub20,de01}.
For instance, Ezawa et al.~\cite{ez97} found a decrease in the iron abundance 
of AWM7 from $\simeq0.5$ solar in the center to $\simeq0.2$ solar at 
a distance of $\simeq500 Kpc$.
These measurements provide strong support for 
the supernova (SN) scenario as heating source for the ICM.

The relative abundance of the heavy elements 
can then be used to constrain the enrichment mechanism of the ICM and the 
energy input from SNe. Analysis of the spatial distribution of metallicity
gradients is also important for discriminating among the proposed
enrichment scenarios \cite{dub20}.
Analytical estimates of the energy input can be substantially different 
according 
to the assumed spatial distribution of the ICM metals and the
efficiency of transfering the kinetic energy released in a SN explosion to the 
ambient gas \cite{kr00}.
Hydrodynamical simulations of galaxy clusters have the advantage over
analytical methods that they take into account self-consistently the dynamical 
evolution of the gas.
The implementation of a metal enrichment model
for the ICM in hydrodynamical simulations is therefore important in order to 
investigate the ICM metal evolution.
Chemical evolution in hydrodynamical simulations has already been 
considered in a variety of contexts 
\cite{ca98,kr00,mo01,li02,ag01}.

In this paper I present results from hydrodynamical SPH
simulations of galaxy clusters, that have been used to investigate 
the dependence of the final iron abundance 
 on a number of model parameters that control the ICM metal enrichment.
This is in done in order to obtain for the simulated clusters a final ICM distribution
which can simultaneously fit a set of observational constraints, such as the 
observed iron abundances and the luminosity-temperature 
relation.

\section{Simulations and results }

 Hydrodynamical TREESPH simulations 
have been performed in physical coordinates for four test clusters.  
A detailed description of the  procedure can be found in \cite{val02}. 
The cosmological model is a flat CDM model, with vacuum energy density
$\oml=0.7$, matter density parameter $\omm=0.3$ and Hubble constant $h=0.7
=H_0/100 Km sec^{-1} Mpc^{-1}$. $\Omega_b=0.015h^{-2}$ is the value of 
the baryonic density. 
The four test clusters have decreasing virial mass and labels L00, L39, L40 
and L119. The virial temperatures range from $\sim 6 KeV$ (L00) down to 
$\sim 1KeV$ (L119).
The simulations have a number of gas particles $N_g \simeq 22,600$. 
For L119 high resolution (H) runs have $N_g\simeq 70,000$. 
The cooling function of the gas particles depends also on the gas metallicity, 
and cold gas particles are subject to star formation.
Once a star particle is created it will release energy into the
 surrounding gas through SN explosions of type II and Ia.
The feedback energy ($10^{51} erg$) is returned to the 
nearest neighbor  gas particles  of the star particle, according to its
lifetime and IMF~\cite{ar87}.  
SN explosions also inject enriched material into the ICM,
thus increasing its metallicity with time. 
The adopted SNIa yields are those of Iwamoto et al.~\cite{iw99} (model W7).
The yields of type II SN are those of Woosley \& Weaver~\cite{woo95} (model B).
For each star particle the ejected masses of metals are calculated at each 
time step  and distributed among the gas neighbors ($\sim 32$) of the star 
particle. 
For the smoothing spline a  common
choice is the SPH kernel \cite{mo01,li02}. 
The profiles shown in the Figure are for a uniform deposition of metals.

 The projected emission  weighted temperature profiles
are plotted as a function of the rescaled radius in panel (a). Data points 
are the
mean error-weighted temperature profiles of 11 cooling flow clusters from 
De Grandi \& Molendi ~\cite{de02}. 
Each smoothed profile has been rescaled to match the last data point.
There is a remarkable agreement of the simulated profiles with data, the 
only important exception being for the two innermost bins.
The increase
of the cluster temperature at the centre is a consequence of the entropy
conservation during the galaxy formation and the subsequent removal of
low-entropy gas \cite{wu02}.
A possible explanation for this discrepancy with the data lies in the fact that 
the temperatures 
being measured are spectral temperatures ~\cite{ma01}.

The projected metallicity profiles are displayed as a function of
distance in panel (b). The data points are the mean profile from the nine cooling
flow clusters of De Grandi \& Molendi ~\cite{de01}. The iron profiles
of L00 and L39 are the ones which are in better agreement with data.
The overall shape of L40 is similar to that of the other two
runs, but has a lower amplitude. The systematic difference between the profiles
of runs L39 and L40 is mostly due to the different dynamical histories
of the two clusters. Therefore, there are uncertainties in the final
profiles which can be as high as $\sim 50 \%$ and are related to to the 
cluster dynamical evolution. 
At outer radii all of the profiles show a radial decay steeper than
that of data points, which in fact can be considered to have an almost constant
profile. It appears very difficult to modify the model parameters of the 
simulations in such a way that the simulated profiles match the data 
points at the outer radii, without also increasing central abundances.

The iron profile of L119H is inconsistent with the data points, the iron
abundances being smaller at all radii. 
This is the coldest of the four test clusters ($T_X\sim 1KeV$). 
The averaged observed profile is that of 9 cooling flow clusters,
with minimum temperatures $\simgt 4 KeV$.
Without measured profiles for cool clusters it is therefore difficult 
to put observational constraints on the different model parameters 
from the simulated profiles.

Global values $A_{Fe}=M_{Fe}/M_H$ of the iron abundances for the simulated 
clusters (open squares) are compared in panel (c) with the 
estimated values (filled squares) for the nearby cluster sample of 
Matsumoto et al. \cite{ma20}. 
There is good agreement with data,
but L119H has  $A_{Fe} \simeq 0.2$, which is about a factor $\sim$ two smaller 
than that expected from extrapolating the sample average below the minimum
 cluster sample temperature ($\simeq 2KeV$ ).
For cool clusters  there is the clear tendency in the simulations to produce 
 a smaller amount of iron than that inferred from observations.

\begin{figure}
\hspace{2truecm}
\psfig{figure=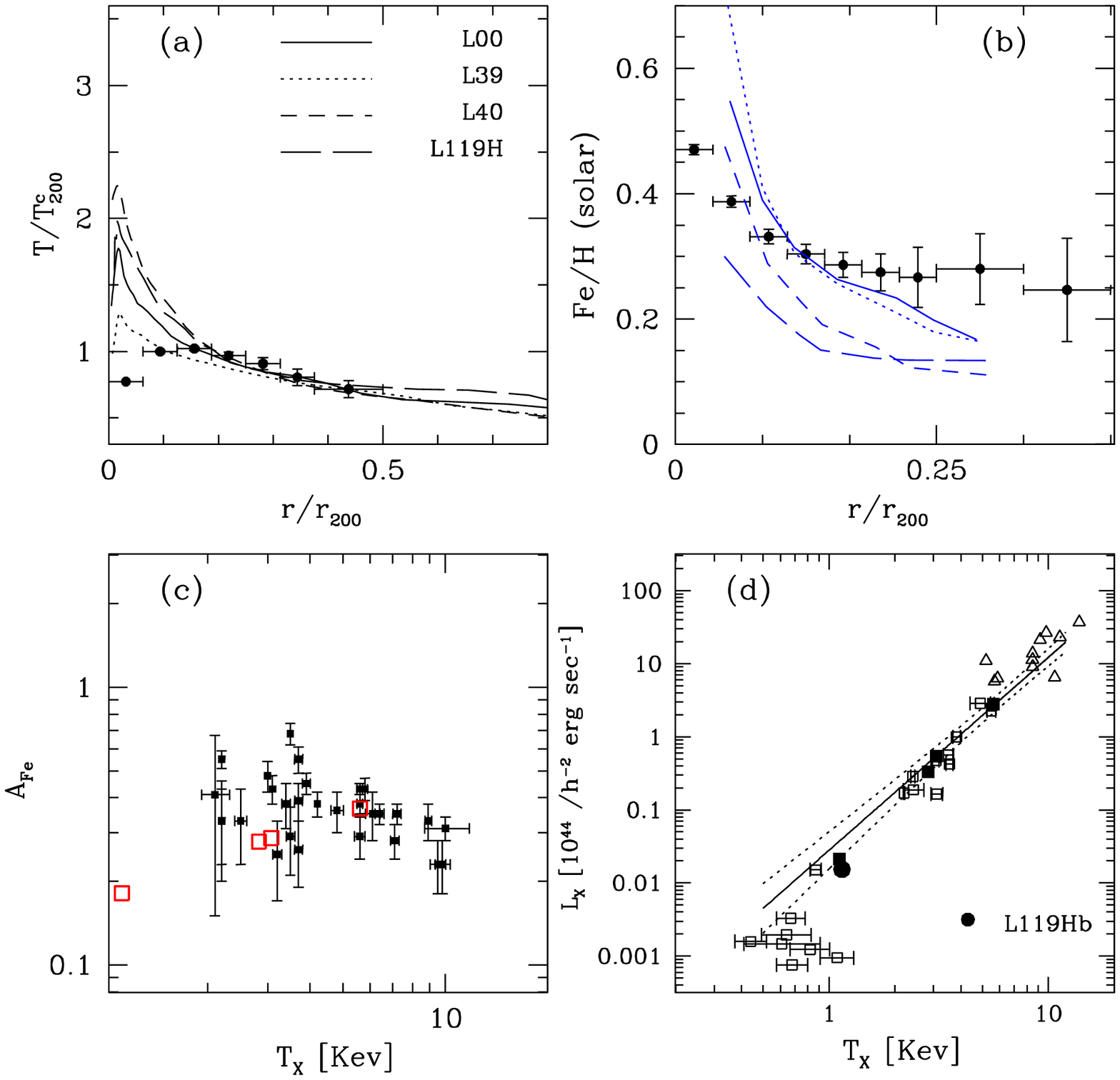,width=12.truecm,clip=}
%\caption{Figure 1}
\label{fig:prof}
\end{figure}

Finally, the values of the bolometric X-ray luminosity are shown in 
panel (d) as a function of the cluster temperature. Mass-weighted temperatures
have been used as unbiased estimators of the spectral temperatures
\cite{ma01}.
Data points (open squares) are those of Fig. 11 of Tozzi \& Norman ~\cite{to01}.
 For the sake of clarity, not all of the points of their Figure are plotted 
in the panel. For a consistent comparison with data, a central 
region of size $50 h^{-1} Kpc$ has been excised ~\cite{mar98}, in 
order to remove the contribution to $L_X$ of the central cooling flow.
The continuous line is 
the best-fit ~\cite{mar98} $L_X=3.11\cdot 10^{44} h^{-2} (T/6KeV)^{2.64}$.
The dashed lines are the $68\%$ confidence intervals. 
  The $L_X$ of the simulations are in excellent
agreement with the data over the entire range of temperatures.
An additional run (L119Hb) has been performed for the cluster with the lowest
temperature, the only difference with respect to L119H being a SN feedback 
energy of $10^{50}$ erg for both SNII and Ia. 
 As can be seen, the final $L_X$  of the run 
L119Hb is very similar to that of L119H. This demonstrates that 
final X-ray luminosities of the simulations are not sensitive to the 
amount of SN feedback energy that has heated the ICM. In fact, a 
simulation run with  zero SN energy being returned to the ICM 
yields very similar results. These findings are particularly relevant
in connection with the recent proposal \cite{br00} that the X-ray properties
of the ICM are driven by the efficiency of galaxy formation, 
rather than by heating due to non-gravitational processes. 
To summarize, the metal enrichment of the ICM can be modeled in hydrodynamical
SPH simulations with a number of model parameters. The results show final
iron abundance gradients in broad agreement with the data. Moreover, the final
X-ray properties of the simulated clusters are not sensitive to the amount 
of SN energy that has heated the ICM. These findings support a scenario where 
the ICM evolution
of cool clusters is dominated by cooling and galaxy formation.

\end{document}